\documentstyle[epsf]{mn}           
\def \src {Cygnus X-1}

\def\OSO {\it OSO-7}
\def\Cop {\it Copernicus}
\def\SAS {\it SAS-3}
\def\MPC {{\it Einstein} MPC}
\def \MPCSS {{\it Einstein} MPC, SSS}
\def\Ten {\it TENMA}
\def\Ginga {{\it GINGA} LAC}
\def\BB {\it BBXRT}
\def\Ros{{\it ROSAT} PSPC}
\def\ME {{\it EXOSAT} ME}
\def\As {{\it ASCA} GIS}
\def \sun {\hbox {$\odot$}}
\def \degmark{^\circ}
\def \nh {$N_{\rm H}$}

\def \hcm {\hbox {\ifmmode $ H atom cm$^{-2}\else H atom cm$^{-2}$\fi}}

\begin{document}
   \title[Distribution of X-ray dips]
         {The Distribution of X-ray Dips with Orbital Phase in Cygnus\thinspace X-1}

   \author[M. Ba\l uci\'nska-Church et al.]
      {M. ~Ba\l uci\'nska-Church$^1$, M. J. ~Church$^1$,
       P. A. ~Charles$^2$, 
\newauthor
        F. Nagase$^3$, J. ~LaSala$^4$ and R. Barnard$^1$\\
	$^1$University of Birmingham, School of Physics and Astronomy,
              Birmingham, B15 2TT, UK\\
	e-mail: mbc@star.sr.bham.ac.uk, mjc@star.sr.bham.ac.uk\\
	$^2$University of Oxford, Department of Astrophysics,
	      Keble Road, Oxford OX1 3RH, UK\\
	e-mail: p.charles1@physics.oxford.ac.uk\\
	$^3$Institute of Space and Astronautical Science,
	       Yoshinodai 3-1-1, Sagamihara, Kanagawa 229-8510, Japan\\
	e-mail:nagase@astro.isas.ac.jp\\
        $^4$University of Southern Maine, Physics Department, 
	       Portland, Maine 04104-9300\\
	e-mail: lasala@lasala.phy.usm.maine.edu}
\date{Accepted. Received}

\maketitle

\begin{abstract}
We present results of a comprehensive study of the distribution of
absorption dips with orbital phase in Cygnus\thinspace X-1. 
Firstly, the distribution was obtained using
archival data from all major X-ray observatories and corrected for the 
selection effect that phase zero (superior conjunction of the black hole)
has been preferentially observed. Dip occurrence was seen to vary
strongly with orbital phase $\phi$, with a peak at $\phi$ $\sim$0.95,
i.e. was not symmetric about phase zero. Secondly, the {\it RXTE} ASM has
provided continuous coverage of the Low State of Cygnus\thinspace X-1 since Sept.
1996, and we have selected dip data based on increases in hardness
ratio. The distribution, with much increased numbers of dip events,
confirms that the peak is at $\phi$ $\sim$ 0.95, and we report the discovery
of a second peak at $\phi$ $\sim$ 0.6. We attribute this peak to absorption
in an accretion stream from the companion star HDE\thinspace 226868. We
have estimated the ionization parameter $\xi$ at different positions
showing that radiative acceleration of the wind is suppressed by 
photoionization in particular regions in the binary system. To obtain 
the variation of column density with phase, we make estimates of neutral
wind density for the extreme cases that acceleration of the wind is
totally suppressed, or not suppressed at all.
An accurate description will lie between these extremes. In each case, a
strong variation of column density with orbital phase resulted, similar
to the variation of dip occurrence. This provides evidence that formation
of the blobs in the wind which lead to absorption dips depends on the
density of the neutral component in the wind, suggesting possible
mechanisms for blob growth.
\end{abstract}


\section{Introduction}

\subsection{Cygnus\thinspace X-1}

\src\ is well known as a black hole binary, with a mass of the compact
object in the range 4.8 -- 14.7~M$_{\sun}$ (Herrero et al. 1995). It is a 
High Mass X-ray Binary, and a member
of the subclass of Supergiant X-ray Binaries (SXBs) consisting of a neutron star
or black hole and an OB supergiant companion. When the compact object is a 
neutron star, it orbits within a few stellar radii of the OB star, and so is
deeply embedded in the stellar wind. For the larger  mass of the compact object 
in Cygnus\thinspace X-1, the orbit is not so close. 
\src\ is extremely variable over a large range of timescales.
First, there are transitions between the low and high luminosity states; \src\
spends most of its time in the low luminosity state, but in May 1996, it underwent 
a transition from the Low State to an intermediate or Soft State (Cui et
al. 1996), returning
after three months to the Low State. On short timescales it is also
very variable, due to the well-known rapid aperiodic variability 
or flickering, the nature of which is not well understood. The X-ray spectrum of
\src\ in the Low State can be described by a hard underlying power law
plus a reflection component (Done et al. 1992) and a weak soft excess 
(Ba\l uci\'nska and Hasinger 1991) which has been identified with blackbody emission
from the accretion disk (Ba\l uci\'nska-Church et al. 1995). In the Soft State, the
spectrum is dominated by a strong thermal component produced by enhanced emission
from the accretion disk (Dotani et al. 1996; Cui et al. 1997).

\subsection{X-ray Dip Properties}

X-ray dips are also observed in Cygnus\thinspace X-1, which usually last several minutes
but have been up to 8 hrs in length. During the dips, there is a spectral 
hardening and the K absorption edge of iron may be seen showing that they are 
due to photoelectric absorption (e.g. Kitamoto et al. 1984). Spectral fitting of 
the X-ray spectra of \src\ has shown that the column density can increase from 
the non-dip value of $\rm {\sim 6.0\times 10^{21}}$ H atom cm$^{-2}$ to
$\rm {\sim 1\times 10^{23}}$ H atom cm $^{-2}$, in dipping (Ba\l uci\'nska-Church 
et al. 1997; Kitamoto et al. 1984). Spectral fitting of dip data, for
example from {\it ASCA} is consistent with neutral absorber although
discriminating between cold and warm absorber was not possible (Ba\l
uci\'nska-Church et al. 1997). Kitamoto et al. (1984) from analysis of 
a high quality {\it Tenma} spectrum of a dip, found an Fe edge implying
an ionization state $<$ Fe V. It was realised at an early stage
that dipping tends to occur at about phase zero of the 5.6~d orbital cycle,
i.e. near to superior conjunction of the black hole, but dipping was also
seen, for example,
at $\phi \sim$0.71 (Remillard \& Canizares 1984). Possible causes of dipping
that have been suggested (see Remillard \& Canizares 1984) are: 1)
absorption asociated with Roche lobe overflow, matter having however, to
be far out of the orbital plane; 2) absorption taking place in the stream flowing 
from the companion towards the compact object inferred from He II $\lambda$ 4686 
measurements (Bolton 1975; Treves et al. 1980); 
3) absorption in the wind of the companion 4) absorption in blobs in the
wind of the companion (Kitamoto et al. 1984). Kitamoto et al. used the
duration of short dips to estimate the size of an absorbing cloud as 
$\sim$ 10$^9$ cm, i.e. a relatively small region, and hence associated the
absorber with clouds or ``blobbiness'' in the stellar wind of the
companion. The present position is that the physical state and origin of
the absorber is not at all well understood.

This contrasts with X-ray dipping in low mass X-ray binaries in which it
is generally accepted that dipping is due to absorption in the bulge in the outer
accretion disc where the accretion flow from the companion impacts 
(White \& Swank 1982). Spectral evolution in dipping could not be
explained in terms of absorption of a single emission component,
since in particular sources, the spectrum may become harder, remain energy 
independent or even become softer during dipping. However, this behaviour has
been explained by assuming two emission regions: point-like blackbody
emission from the surface of the neutron star plus extended Comptonized
emission from the accretion disk corona (Church \& Ba\l uci\'nska-Church 1995;
Church et al. 1997, 1998a, 1998b). This model is able to explain the
varied and complex spectral evolution in dipping in different sources.
The presence of X-ray eclipses in 
XBT\thinspace 0748-676 shows that deepest dipping occurs at orbital phase 
$\sim$0.9 consistent with the position of impact on the disk of an accretion 
flow trailing sideways from the inner Lagrangian point in the binary frame. Thus, 
dipping is much better understood in LMXBs than in \src.

\subsection{Supergiant Stellar Winds}

SXBs in general have strong stellar winds and exhibit strong
orbital-related decreases in X-ray intensity due to absorption in the wind.
In an {\it EXOSAT} observation of the archetypal eclipsing SXB, 
X\thinspace 1700-371, it was found that large, smooth increases in column density 
took place between orbital phases 0.8 and 1.2 (Haberl, White \& Kallman
1989; hereafter HWK89). This observation was useful since a full orbital
cycle of 3.41 day was covered. In X\thinspace 1700-371, the binary
separation is the smallest known, with the compact
object orbiting the primary star at 1.4 stellar radii (van Genderen 1977),
leading to the dramatic changes in absorption with orbital phase.
The increases in column density could be well modelled by absorption in a stellar wind obeying 
a CAK velocity law (Castor, Abbott \& Klein 1975). An additional sharp 
increase in column density at phase $\sim$0.6 could be well modelled as a 
gas stream originating on the companion, possibly on a tidal bulge 
(HWK89). 
In the companion of Cygnus\thinspace X-1,
HDE\thinspace 226868, the radial velocity curve of the He II $\lambda$ 
4686 emission line is shifted by about 120$\degmark$ with respect to 
absorption in the companion, also indicating the presence of an accretion 
stream (Hutchings et al. 1973). In the case of \src , a strong decrease in X-ray
intensity with orbital phase has not been seen; however,
Kitamoto et al. (1990) showed evidence for a dependence of column density
of the quiescent (non-dip) spectra on orbital phase, but with column density
increasing from $\rm {\sim 6\times 10^{21}}$ H atom cm$^{-2}$
to only $\rm {2\times 10^{22}}$ H atom cm$^{-2}$, at least one order 
of magnitude less than in X\thinspace 1700-371.
Apart from this, the only absorption events seen are the X-ray dips.
There has been no systematic study of the phase of dipping, and in this
work we present such a systematic study, and compare dipping in \src\ with 
absorption effects in other SXBs.

\subsection{Survey of X-ray Dips}

It has not previously been possible to
make a survey of the distribution of dips with orbital phase because
of uncertainties in the available ephemeris.  The ephemeris previously
available of Gies \& Bolton (1982) gave an orbital period of
5.59974 $\pm $0.00008 days, and this precision implies an uncertainty
in phase of $\pm$0.02 cycle at the present time. However, Ninkov, Walker
\& Yang (1987) presented evidence for a period increase with time,
and phases calculated using their ephemeris now differ by 0.5 from Gies \& Bolton
values, so that phase becomes completely indeterminate at the present time.
In response to this, a new definitive ephemeris has recently been
produced based on high quality radial velocity data on HDE\thinspace 226868,
by LaSala et al. (1998). This work
concludes that there is no evidence for an evolving period, and the
accuracy of the period determination of P = 5.5998 $\pm$ 0.0001 days in the new
ephemeris allows accurate retrospective determination of orbital phase back 
many years from the present epoch; in fact, 85 years are required before an uncertainty 
of 0.1 cycle accumulates. The epoch of the new ephemeris is given for
the superior conjunction of the black hole, i.e. $\phi$ = 0 
with the companion between the observer and the black hole, and thus
corresponds to the orientation in which X-ray dips are thought 
to take place.  

\begin{table}
\caption[]{The survey of X-ray dips}
\begin{flushleft}
\begin{tabular}{rllllll}
\noalign{\hrule\smallskip}
Date  &&&Observatory &  Ref.  \\
\noalign{\hrule\smallskip}
 1972 &Nov &6$^{th}$    &      \Cop  & 1 \\
 1972 &Dec &31$^{st}$    &      \OSO   & 2 \\
 1973 &May &16$^{th}$   &      \Cop  & 1 \\
 1973 &Sep &4$^{th}$  &      \Cop  & 1 \\
 1973 &Oct &30$^{th}$   &      \Cop  & 1 \\
 1973 &Nov &5$^{th}$   &      \Cop  & 1 \\
 1976 &Aug &4$^{th}$  &      \SAS  & 3 \\
 1976 &Oct &10$^{th}$   &      \SAS  & 3 \\
 1977 &Aug &10$^{th}$   &      \SAS  & 3 \\
 1978 &Nov &21${st}$  &      \MPC  & 4 \\
 1979 &Apr &15$^{th}$   &      \MPC  & 4 \\
 1979 &May &9$^{th}$   &      \MPCSS  & 4,5 \\
 1979 &May &13$^{th}$   &      \MPCSS  & 4,5 \\
 1979 &May &30$^{th}$   &      \MPC  & 4 \\
 1979 &Oct &12$^{th}$   &      \MPC  & 4 \\
 1979 &Oct &18$^{th}$   &      \MPC  & 4 \\
 1983 &Sep &9$^{th}$   &      \Ten  & 6 \\
 1984 &Jul &8$^{th}$  &      \ME   & 7 \\
 1985 &Oct &15$^{th}$   &      \ME   & 7 \\
 1987 &Aug &5$^{th}$    &     \Ginga & 8  \\
 1990 &May &9$^{th}$   &      \Ginga & 8  \\
 1990 &Dec &6$^{th}$  &      \BB   & 9 \\
 1991 &Apr &18$^{th}$   &      \Ros  & 10 \\
 1991 &Jun &6$^{th}$    &      \Ginga & 8 \\
 1994 &Nov &23$^{rd}$   &      \As   & 11 \\
 1995 &May &9$^{th}$   &      \As   & 12 \\

\noalign {\smallskip}
\noalign {\hrule\smallskip}
\noalign {\smallskip}

\end{tabular}
\end{flushleft}
{References: $^1$ Mason {\it et al.} 1974; $^2$ Li \& Clark 1974;
$^3$ Remillard \& Canizares 1984; $^4$ Ba\l uci\'nska 1988; $^5$ Pravdo
{\it et al.} 1980; $^6$~Kitamoto {\it et al.} 1984; $^7$ Ba\l uci\'nska
\& Hasinger 1991; $^8$ {\it present work}; 
$^9$ Marshall {\it et al.} 1993; $^{10}$ Ba\l ucinska-Church
{\it et al.} 1995; $^{11}$ Ebisawa {\it et al.} 1996; $^{12}$ Ba\l uci\'nska-Church
{\it et al.} 1997}
\end{table}

We are thus able to present a survey of the
distribution of dipping with orbital phase, and this is based on two sources
of data. First, a survey has been made using a large body of archival data  
from {\it Copernicus}, {\it Einstein}, {\it EXOSAT}, 
{\it TENMA}, {\it GINGA}, {\it ROSAT} and {\it ASCA}. Secondly, we have used data 
from the {\it Rossi-XTE} All Sky Monitor (ASM) obtained between September 1996
and the present.

This distribution shows a smooth variation with orbital phase which is
strikingly similar to the observed variation of column density with
phase in the HMXB X\thinspace 1700-371 (HWK89), superimposed on the
smooth variation is evidence for a stream. Consequently, we attempt
simple modelling of the wind of HDE\thinspace 226868 assuming that
most of the wind is highly ionized by the X-ray flux from the black
hole, and from the density of the neutral component of the wind, derive
the variation of column density along lines-of-sight with orbital phase.

\subsection{Wind-Driving Mechanisms}

The SXBs, including Cygnus\thinspace X-1 of spectral type O9.7Iab,
consist of OB supergiants and a compact object, and the
massive companion has a strong stellar wind as do isolated OB stars.
The stellar wind
in such systems is driven by the radiation pressure of UV photons
on the gas. Calculations of the radiation force have been made by Lucy
\& Solomon (1970) based on resonance lines from a few elements, by
Castor, Abbott and Klein (CAK; 1975) based on the combined effect of
weak lines, and by Abbott (1982) who concluded that the correct approach
uses the strong lines of many elements. However, in the case of binary
systems, the X-ray luminosity L may result in photoionization
which will suppress the radiative driving force. For high L,
radiative acceleration will be totally suppressed on the side of the
companion
exposed to X-rays. A two-dimensional hydrodynamic coded was used by
Blondin et al. (1990) to include effects such as radiative driving of
the wind, suppression by photoionization, X-ray heating, radiative
cooling and gravitational and rotational forces in the binary system.
By application of this code to high X-ray luminosity systems such as
Cen\thinspace X-3 and SMC\thinspace X-1 Blondin (1994) produced
two-dimensional maps of particle density for L between
$\rm {10^{36}}$ and $\rm {10^{38}}$ erg s$^{-1}$ showing dramatically
the suppression of the normal wind and also demonstrated the production
of a photoionization wake where the normal wind meets a stalled wind, and
of a shadow wind originating in the X-ray shadow moving sideways in the
binary system towards the compact object on the upstream side.
The simulations of particle
density are then used to calculate column density values along
lines-of-sight to the X-ray source on the assumption that material
having
ionization parameter $\xi$ $<$ 2000 will contribute to photoelectric
absorption.

In the case of Cyg\thinspace X-1, we take a simple, approximate approach
of calculating the density of the recombined component of a wind that is
assumed to be
fully ionized and from this derive column densities as a function of
orbital phase for comparison with the dip distribution.

\begin{figure}
\epsfxsize=80mm
\begin{center}
\leavevmode\epsffile{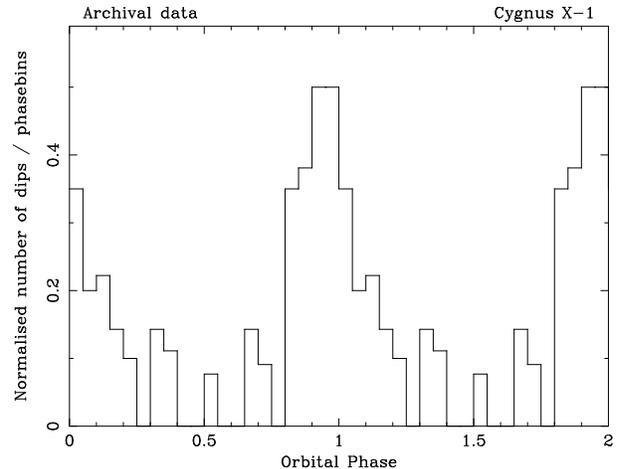}
\end{center}
\caption{Frequency of dip occurrence as a function of orbital phase,
normalized by dividing each value by the number of times each phasebin
was observed. The peak represents dipping activity in 10 separate
orbital
cycles at phase 0.90--0.95 from a total of 20 observations (and dipping
in 11 of 22 observation of phase 0.95--1.0 \label{fig1}}
\end{figure}

\section{Observations}
\label{sec:observations}

\subsection{Archival data}

\begin{figure*}
\epsfxsize=120mm
\begin{center}
\leavevmode\epsffile{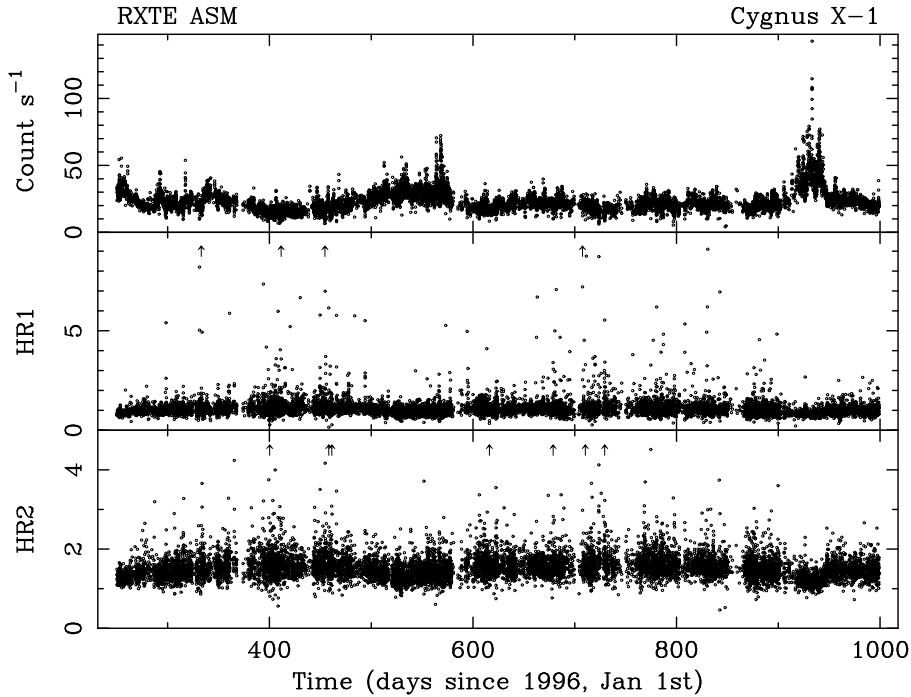}
\end{center}
\caption{ASM light curve and hardness ratios HR1 and HR2
based on the standard energy bands (see text). To emphasise the
constancy of the non-dipping hardness ratios, several points with
large hardness ratios have been omitted and are indicated by arrows
\label{fig2}}
\end{figure*}

For the first part of our analysis, we have included the data from a wide
range of X-ray missions as shown in Table 1, which gives the date of
the start of each observation, the observatory mission, and the authors.
In using data published by other authors, we have used tables and figures 
showing the times of dipping from the literature. In other cases, data was 
analysed by ourselves, i.e. {\it Einstein} MPC: Ba\l uci\'nska (1988),
{\it EXOSAT}: Ba\l uci\'nska and Hasinger (1992),
{\it GINGA}: {\it Present work}, 
{\it ROSAT}: Ba\l uci\'nska-Church et al. (1995), and {\it ASCA}: 
Ba\l uci\'nska-Church et al. (1997).

\subsection{{\it Rossi-XTE ASM} data}

The All-Sky Monitor (ASM; Levine et al. 1996) on board {\it Rossi-XTE} 
(Bradt, Rothschild \& Swank 1993) scans the sky in 3 energy bands: 1.5{\bf--}3, 
3{\bf--}5 and 5{\bf--}12~keV with $\sim$90~s exposure. Any source is scanned 
5{\bf--}20 times per day. We have extracted data from the archive, in these 
3 energy bands, including data from 
September 6th, 1996 until the present (October 1998), i.e. including data
labelled as days 250 - 998 from the start of ASM operation. The
start time was chosen so as to exclude the period between 1996 May to 
August when \src\ was in the Soft State; thus all of the results presented
relate to the Low State of the source.

\section{Analysis and Results}
\subsection {The Historical Data}

In all cases of archival data, orbital phases were calculated using
the new ephemeris of LaSala et al. (1998). First, the data were
sorted into 100 phasebins, and a count was added to each bin in which
dipping activity was observed.
In \src\, a succession of narrow X-ray dips usually occurs, and so we
add a single count to a given phasebin when dipping is seen, not a count
for each narrow dip. 

There would be a strong selection
effect due to the fact that coverage of phases has deliberately
concentrated on phases close to zero. To compensate for this, we also
made a count of the number of times each phasebin was observed, and 
the histogram of dip frequency {\it versus} phase was
normalized by dividing the dip count by the phasebin count.
A selection effect may remain, as data not analysed by ourselves,
was selected by the various authors as suitable for publication.
Finally, to improve statistics, phasebins were grouped together
in fives to give 20 bins per orbital cycle, and this is 
shown in Fig.~1.

It can be seen that the distribution peaks at about phase 0.95
with a full width at half maximum of $\sim$0.25. This effect, i.e.
the peak of the distribution being offset from phase zero, could also be seen
in individual datasets; for example, Remillard and Canizares (1984) noted that
short dips tended to occur before phase zero, whereas one 8 hr dip 
was centred on phase zero. This will be discussed fully after the next section 
on results from the {\it RXTE} ASM.

\subsection {{\it RXTE} All Sky Monitor Data}

Data were extracted from  
the ASM archive and stored as entries in a file, each entry corresponding
to a single 90~s observation. Each entry consisted of the time, and the
count rate in each of the 3 standard bands: 1.5{\bf--}3~keV, 
3{\bf--}5~keV and 5{\bf--}12~keV. From these, hardness ratios were 
constructed using the ratio of count rates in the bands 3{\bf--}5~keV and 
1.5{\bf--}3~keV designated HR1, and the ratio of count rates in the bands
5{\bf--}12~keV and 3{\bf--}5~keV, designated HR2. 
The lightcurve in the total energy band of the ASM, together with
HR1 and HR2, is shown in Fig.~2. A brief period of enhanced intensity
can be seen at $\sim$ day 920; however, this data is excluded by our
procedure of selecting data dip data using hardness ratios (below). 
Although the count rate varies between
15 and $\sim$140 count s$^{-1}$, mostly due to flaring activity,
the hardness ratios HR1 and HR2 were very stable, with mean values
1.10$\pm 0.77$
and 1.52$\pm 0.35$, respectively and were unaffected by flaring. 
The points with hardness ratio significantly larger than these means are 
dip data, for which HR1 can increase to 40 and HR2 can increase
to 10.

Dip events can be seen in the light curve, but are more obvious in HR1 and
HR2. The data folded on the orbital period are shown in Fig.~3.
There is a clear anti-correlation between 
the hardness ratios and the count rate, with a reduction in count 
rate occuring just before phase zero, and associated hardening of the 
spectrum showing that it is due to absorption. Larger increases of hardness
ratio in the lower energy bands would be expected for simple absorption;
however, as partial covering takes place, the decrease in intensity at low
energies is reduced by the presence of the uncovered part of the emission.
The data can also be plotted 
as a colour-colour diagram, i.e. as HR2 against HR1, and this is shown in 
Fig.~4. We next discuss how spectral simulations were used to elucidate
the behaviour seen in Fig.~4, and how selection of dip data was made.

\subsubsection {Flux calibration of the ASM and simulations of dipping in the 
ASM colour-colour diagram}

\begin{figure}
\epsfxsize=80mm
\begin{center}
\leavevmode\epsffile{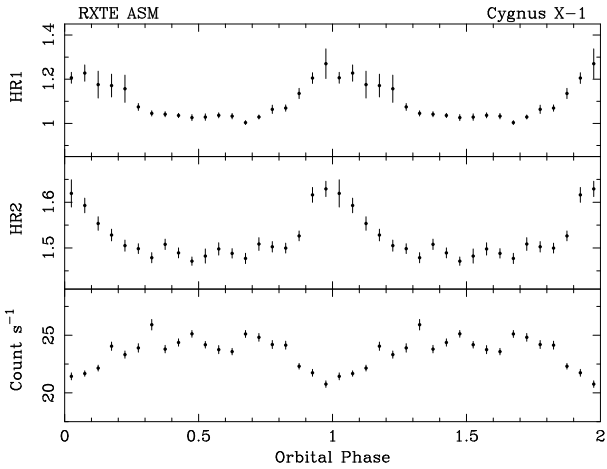}
\end{center}
\caption{ASM data on Cygnus\thinspace X-1 sorted into 20 phasebins by
folding on the
orbital period: total count rate (lower panel) and hardness
ratios HR1 and HR2
\label{fig3}}
\end{figure}

To understand the spectral changes taking place in the data plotted in
Fig.~4, we have simulated non-dip and dip spectra and obtained the fluxes
in the 3 ASM bands. These were then converted to count rates in the 3
bands, and from these, the two hardness ratios HR1 and HR2 were derived. 
To calibrate the flux/count rate relations in each of the bands, not
having an instrument response for the ASM, we have used ASM data from the
Crab Nebula.
The X-ray flux of the Crab in each of the standard bands was
calculated and compared with the mean value of count rate in each band integrated 
over a long period. This gives calibration factors 
of $\rm {\sim 3\times 10^{-10}}$ erg cm$^{-2}$ s$^{-1}$ per count
s$^{-1}$ for each of the 3 
energy bands. This calibration makes the assumption that the 
spectral shape of the source does not differ markedly from that of the Crab 
Nebula over the energy range of the ASM.

\begin{figure}
\epsfxsize=80mm
\begin{center}
\leavevmode\epsffile{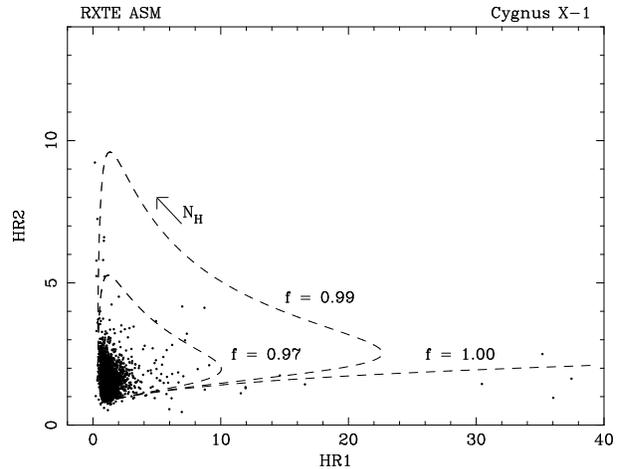}
\end{center}
\caption{Colour-colour diagram of ASM data including non-dip and dip
points. Superimposed are simulation results based on typical spectral
parameters and a partial covering model for dipping. The dashed lines
show tracks at constant covering fraction {\it f} with $\rm {N_H}$
increasing anticlockwise
\label{fig4}}
\end{figure}

Next, we simulated dipping in \src\ for a wide range of spectral
parameter values as
follows. Spectral fitting of the source in the Low State
showed that the contribution of the blackbody soft excess emission from 
the accretion disk is small, and so the source may be 
modelled with a power law spectrum which is partially covered during dipping 
(Ba\l uci\'nska-Church et al. 1997). Good fits were obtained using neutral
absorber with solar abundances (cross-sections of Morrison \& McCammom
1983).
Dip spectra are normally made by
intensity selection to accumulate sufficient counts, and so include dip data 
due to a succession of short dips, plus non-dip data between the dips, and
this accounts for the partial covering required in fitting.
The parameters {\it f}, the partial 
covering fraction and the column density \nh\  can cover 
wide ranges of values and, in simulations, we spanned the whole range of this two-dimensional 
space in a set of calculations making small steps in the parameter values, 
calculating the flux in each of the ASM bands, and plotting the results on 
a colour-colour diagram of HR2 {\it versus} HR1. These simulated data show 
all the features of the actual colour-colour diagram of \src\ data obtained 
from the ASM (Fig.~4), and allow us to interpret the real data in different
positions on the plot.

First, the limited number of real dip points moving horizontally in HR1
with little increase in HR2 correspond to dipping with partial covering fraction
{\it f} = 1.0, i.e. simple absorption in which the source region is
completely covered by absorber. Along this horizontal line, \nh\
increases from low values to a maximum of 
$\rm {\sim 25\times 10^{22}}$ H atom cm$^{-2}$. Other positions on the 
colour-colour diagram have {\it f} $<$ 1. For a given {\it f} value, increasing
\nh\ traces a loop on the plot with
column density increasing in an anticlockwise sense. For very high \nh,
the count rate in the highest energy band begins to be affected, and 
points move on a vertical line at small HR1, so that as \nh\
continues to increase, the data points move vertically {\it downwards}.

Based on the above understanding of the spectral changes taking place in
the source revealed in Fig.~4, we selected dip data by taking points with 
a value of HR1 $>$ 2 and HR2 $>$ 2.5. The plots of HR1 and HR2 against time 
were very flat (Fig.~2) allowing this selection method
to be used. 
Kitamoto et al. 
(1990) found evidence for a variation of the {\it non-dip} column density 
with orbital phase from $\rm {\sim 6\times 10^{21}}$ H atom cm$^{-2}$ to 
$\rm {\sim 2\times 10^{21}}$ H atom cm$^{-2}$, which they associated with 
varying \nh\ in the stellar wind. Our simulations show that such 
increases in \nh\ will be excluded from the selection of dip data. 

For each point selected from the total data set in Fig.~4 as dip data,
the orbital phase was calculated 
using the new ephemeris (LaSala et al. 1998) and the distribution is shown 
in Fig.~5. The data are not normalised by the time spent at each phase as 
the observation of orbital phases was almost uniform. In principle, it
might be possible to derive \nh\ values for each data point using
our simulations; however this is made difficult by the fact that any point
on the colour-colour diagram has particular \nh\ and {\it f} values,
and adjacent points can have very different \nh. Although we
cannot produce a plot of \nh\ {\it versus} phase, by plotting HR1
against phase for the selected data, it appears that the points at $\phi$
$\sim$0 have the largest values of HR1 and thus the largest values of column
density.

\begin{figure}
\epsfxsize=80mm
\begin{center}
\leavevmode\epsffile{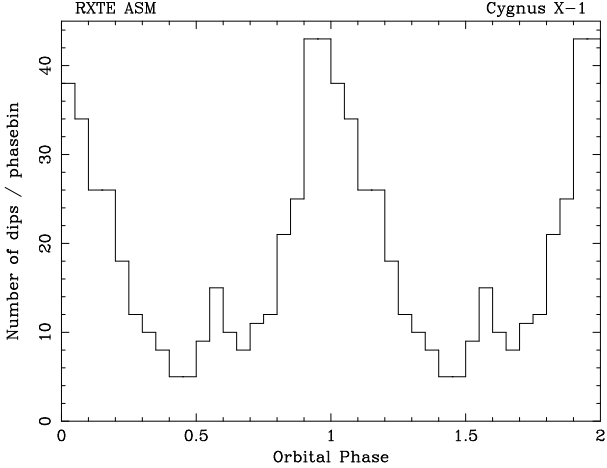}
\end{center}
\caption{Distribution of X-ray dips with phase from the ASM
\label{fig5}}
\end{figure}

\subsubsection {The distribution of dipping with phase}

The results shown in Fig.~5 show a similar effect to the historical data,
i.e.
the peak of dipping occurs offest from phase zero at phase $\sim$0.95.
In addition, a peak at phase 0.6 can be seen. The statistics are much improved
compared with the archival data. A total of 12519, 90$\,$s observations from the
ASM were extracted, made at an average of 16.7 observations per day, of which 
about 3\% were dip events giving a total of 379 dip events compared 
with 55 events in the archival data. The two effects seeen
in the distribution against orbital phase are very similar to the effects
seen in SXBs in measurements of column density of the X-ray spectrum as
a function of orbital phase, e.g. in X\thinspace 1700-371 (HWK89). An asymmetry
in \nh\ about phase zero and a peak in \nh\ are seen in this case.
Thus there is evidence that the distribution of dipping relates ultimately
to the properties of the stellar wind and the existence of a stream. It is
thus important to know the variation of neutral wind density with orbital phase in
Cygnus\thinspace X-1, and we have carried out modelling of the stellar wind
as discussed in the next section.

\subsection {Wind modelling}

We wish to derive column densities as a function of orbital phase for 
comparison with our results on the distribution of dipping with phase.
Cygnus\thinspace X-1 is luminous and we may expect suppression of the 
radiative wind driving force in particular spatial regions. Consequently 
we make approximate 
calculations of wind density and ionization parameter $\xi $ to
investigate the extent of the suppression, and then estimate values of
$\rm {N_H}$.

\begin{figure}
\epsfxsize=80mm
\begin{center}
\leavevmode\epsffile{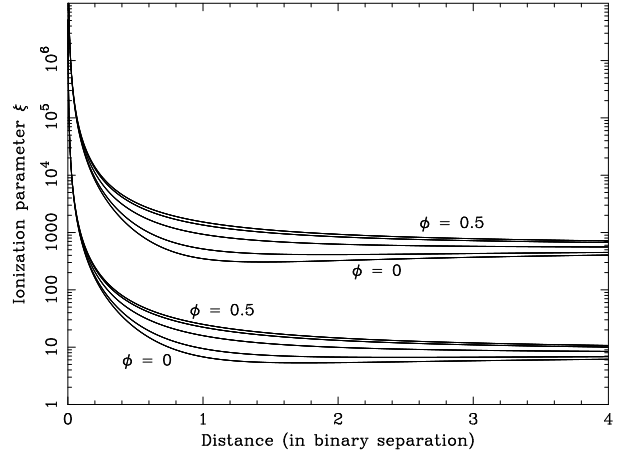}
\end{center}
\caption{Ionization parameter $\xi$ as a function of distance from the
black hole along lines of sight at a range of orbital phases between 0
and 0.5. The two sets of curves represent extreme assumptions: the lower 
curves assume total suppression of the radiatively-driven wind; the upper 
curves assume no suppression. Distance is measured in units of the separation of the
stars
\label{fig6}}
\end{figure}

First we calculate the ionization parameter $\xi$ = $\rm {L/nr^2}$,
where n is the particle density, as a function of distance r
from the black hole to determine the extent of suppression of the wind
driving force. This is done for different orbital phases along lines of sight that 
will be used to integrate particle number density to give column
density, and for simple assumptions about the ionization state of the wind, i.e.
about whether acceleration is suppressed or not. Assuming first that 
the wind is generally highly ionized we calculate the densities 
of neutral atoms $\rm {n_0}$ and of electrons $\rm {n_e}$ 
in a wind which drifts with constant velocity $\sim$ 30 km s$^{-1}$ as
a velocity typical of turbulent motions in the atmosphere of HDE\thinspace 226868 
(Gies \& Bolton 1986). The total particle density in the wind
$\rm {n_w}$ ($\rm {n_w}$ = $\rm {n_0}$ + $\rm {n_e}$) is given by

\vskip - 2mm
\[\rm {n_w = \dot M /4\pi m_H R^2 v}\]

\noindent
where $\rm {\dot M}$ is the mass accretion rate, R is the distance
from the centre of the primary, $\rm {m_H}$ is the mass of the hydrogen atom and 
v is the wind velocity. Spherical symmetry of the wind is
assumed. A simple approach was taken considering only the equilibrium
between production of ions by photoionization of a solar abundance mixture, 
and radiative recombination of H$^+$ ions. This results in a neutral
component of the order of 1\% of the total wind density. We are
neglecting ionization by the primary (Abbott 1982) which will reduce the
neutral fraction substantially further. However, it is only the X-ray
flux which suppresses radiative driving. If the number density of X-ray photons
is $\rm {n_{\gamma}}$, the photoionization rate $\rm {dn_e/dt}$
is $\rm {k\,n_0\,n_{\gamma}}$, and the recombination rate is 
$\rm {\alpha\, n_e^2}$,
where the photoionization rate coefficient k, is related to the
cross-section $\sigma$ via k = $\rm {\sigma \,c}$ (c is the velocity
of light) and $\alpha$ is the recombination coefficient. The photoionization 
cross-sections of Morrison \& McCammon (1983) were used.
From Allen (1973), $\alpha$ = $\rm {3\times 10^{-11}\,Z^2\,T^{-1/2}}$,
having a value $\alpha$ = $\rm {2.0\times 10^{-13}}$ for T = 10$^4$ K.
In equilibrium,

\[\rm{  {n_{\rm e}^2\over n_{\rm 0}} = {\sigma\,c\,n_{\gamma}\over\alpha}}\]

\noindent
The photon number density $\rm {n_{\gamma}}$ is given by 
$\rm {n_{\gamma}}$ = 
$\rm {L/ 4\,\pi\,{\rm r}^2}\,\overline {\rm {E}}$ where $\rm {\overline {E}}$ 
is the mean energy, 
assumed to be 1~keV and L = $\rm {2\times 10^{37}}$ erg s$^{-1}$ typical of the 
Low State. From the above, a simple quadratic equation for $\rm {n_e}$ follows,
allowing $\rm {n_e}$, $\rm {n_0}$, the fractional ionization and $\xi$
to be obtained. Fig.~6 (lower curves) shows that $\xi$ is high close to the black hole, 
but falls to $<$ 100 at distances
greater than 0.3 of the binary separation; thus the Str\"omgren region
where radiative acceleration is suppressed does not extend down to the 
surface of the companion. In view of this, we also derived $\xi$ curves
using a CAK law assuming no ionization of the wind as shown in the upper 
curves of Fig.~6. This assumed the velocity v(R) increases
as $\rm {v_{\infty} (1 - R_*/R)^{\alpha}}$ where $\rm {v_{\infty}}$ = 
2100 km s$^{-1}$, $\rm
{R_*}$ = 17.0 R$_{\sun}$, $\rm {\dot M}$ = $\rm {3.0\times 10^{-6}}$
M$_{\sun}$ yr$^{-1}$ (Herrero et al. 1995) and $\alpha$ is 0.5. 
In this case, $\xi$ is greater than 1000 even at distances of several 
binary separations from the black hole. With these two extreme models
$\rm {N_H}(\phi)$ was obtained by integrating $\rm {n_0}$ along lines 
of sight to the X-ray source for a range of orbital phases $\phi$ between 
0 -- 0.5 for an inclination angle of 35$\degmark$ and a binary separation
of 40.2 R$_{\sun}$ (Herrero et al. 1995).

It can be seen that the two models produce a similar variation with phase. 
The high values of $\rm {N_H}$ from the drift models would be easily
measurable from the X-ray spectra.
The only observational evidence (Kitamoto et al. 1990) shows a 
variation of similar shape between 6 and $\rm {20\times 10^{21}}$
H atom cm$^{-2}$. The correct description
clearly lies between our 2 extremes; we concentrate
on the relation between the dipping and the $\rm {N_H}$ variation.
In each case, the variation is strong (a factor of 6 for the drift model
and a factor of 10 in the other case) suggesting that blob formation,
and thus dip formation, depends on the neutral density component in the
wind.

\begin{figure}
\epsfxsize=80mm
\begin{center}
\leavevmode\epsffile{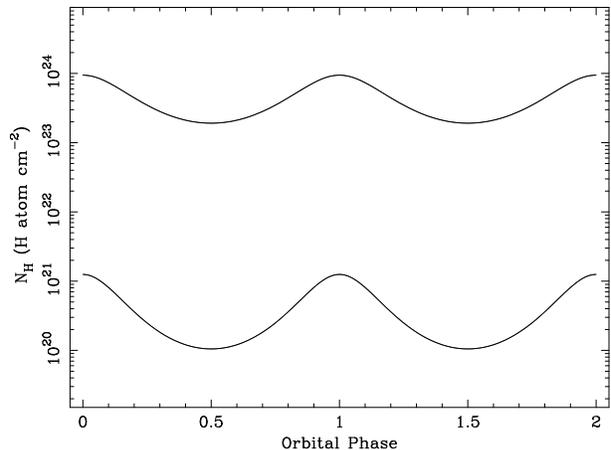}
\end{center}
\caption{Column densities along lines of sight to the black hole for
the same two cases shown in Fig. 6; the upper curves assume total
suppression of the radiatively-driven wind, the lower curves assume
no suppression. 
\label{fig7}}
\end{figure}

\section{Discussion}

Modelling of the stellar wind of HDE\thinspace 226868 should take into
account 2 main effects: radiative driving of the wind and suppression of
the driving force by photoionization due to the X-ray source. To
calculate the column density appropriate to X-ray observations of
Cygnus\thinspace X-1 at various orbital phases would require
three-dimensional calculations of the particle number density in the
wind combined with calculations of the ionization parameter $\xi$.
Suppression of the radiative driving force takes place for $\xi$ $>$ 100 
(Blondin 1994). Detailed modelling using a 2-dimensional hydrodynamic code 
was carried out for the wind density in luminous SXB by Blondin (1994).
In high luminosity systems such as Cen\thinspace X-3 and SMC\thinspace X-1, 
strong effects due to the X-ray source were seen with a Str\" omgren region
about the X-ray source in which radiative driving was suppressed. In
Cen\thinspace X-3, for 
L $>$ $\rm {10^{37}}$ erg s$^{-1}$, this region extended to all of the
X-ray illuminated wind (i.e. not in the shadow of the companion). 
Column densities were obtained by
integrating the particle densities where $\xi$ was $<$ 2000, i.e.
assuming that all material not fully ionized contributes to $\rm {N_H}$.
In our case, the X-ray source is bright even in the Low State 
with L $\sim \rm {2\times 10^{37}}$ erg s$^{-1}$ and it
might be thought that strong suppression of radiative acceleration
takes place. 
However, the binary separation of
40.2 $\rm {R_{\sun}}$ is larger in the black hole binary than the 
19 $\rm {R_{\sun}}$ in Cen\thinspace X-3. Thus the flux of the X-ray
source is reduced by a factor of 4 compared with Cen\thinspace X-3 so
that simple scaling implies total suppression of radiative driving in 
the X-ray illuminated wind only for L $>$ $\rm {4\times
10^{37}}$ erg s$^{-1}$. This depends primarily on $\xi$ and our
simple calculations show that wind with totally suppressed driving force
would result in $\xi$ $>$ 100 only relatively close to the black hole.
Thus it is clear that detailed hydrodynamic simulations of
Cygnus\thinspace X-1
are required to delineate regions where radiative driving is
suppressed. This would also reveal the details of the high density
photoionization wake where normal wind in the X-ray shadow impacts on
stalled wind (Fransson \& Fabian 1980; Blondin 1994), and of the shadow
wind which emerges from the X-ray shadow and may contribute to
absorption (Blondin 1994).

We have shown that there is a strong dependence of the frequency of dipping 
in \src\ with orbital phase, and that this correlates approximately with the 
variation of column density of the neutral component of the stellar wind 
with phase $N_{\rm H}(\phi)$. This suggests that dipping is caused by blobs of 
largely neutral material, the formation of which may
depend simply on the neutral density at any point in the wind.
Our previous spectral fitting has shown that column density is
typically between 2 and $\rm {20\times 10^{22}}$
H atom cm$^{-2}$ in dip spectra (Ba\l ucin\'ska-Church et al. 1997).
For a blob diameter of $\rm {10^9}$ cm (Kitamoto 1984) this gives
densities of $\sim$ $\rm {10^{12}}$ -- $\rm {10^{13}}$ cm$^{-3}$. 
Our estimates for the wind density have maximum values of total wind density between 
a few $\rm {10^9}$ and $\rm {10^{11}}$ cm$^{-3}$ for no suppression of 
radiative driving and total suppression, respectively. A more realistic
typical value might be $\rm {10^{10}}$ cm$^{-3}$. Thus the blob density
is greater than the ambient wind density by factors of 100 - 1000.
In such a higher density region, $\xi$ will be reduced,
so that a high value in the ambient wind of 1000 would be reduced to 
1 - 10, such that the photoionizing effects of the X-ray source are
markedly reduced. One possibility for blob formation is that
neutral material in the wind can act as a nucleus for blob growth, since
in the X-ray shadow of a small blob, photoionization will be reduced
and recombination into the blob will be rapid.
Other possibilities for blob formation also exist. For example,
the interaction of normal wind with stalled wind can lead to a high 
density region trailing behind the compact object (Fransson \& Fabian 
1980; Blondin 1994). In this region, instabilities and density 
enhancements may form. However this would not be exected to produce
dipping having the basic symmetry about $\phi$ $\sim$ 0.5, i.e. the
line of centres, that we see. Similarly this would not be expected to account for
the peak in dipping we see at $\phi$ $\sim$ 0.6 as the high density region 
extends over a large range of angles with respect to the primary
whereas a stream produced by Roche-lobe overflow or tidal enhancement
does not.

In Cygnus\thinspace X-1, for the first time we find definite evidence 
for a stream from X-ray data in the enhanced number of dips at $\phi$
$\sim$ 0.6, and the source appears to be similar to 
X\thinspace 1700-371 in which it was concluded that a stream was
the cause of enhanced absorption at phase 0.6 (HWK89). 
Petterson (1978) showed that HDE\thinspace 226868 is filling its
Roche lobe and will therefore will produce a stream which may be
expected to produce absorption effects at $\phi$ $\sim$0.6  Even if the
star was only close to filling its Roche lobe, Blondin et al. (1991) 
have shown that a  stream will still develop by tidal
enhancement of the stellar wind caused by the compact object.
In the neutron star systems modelled, the 
formation of a stream depends on the binary
separation. However, a clear result of this work was that 
the stream is produced at phase $\phi$ $\sim$ 0.6. In either case,
a stream is produced at a phase similar to that we found here. Blob
formation in a stream would, of course, be easier than in the wind
because the density is already increased over the wind density
reducing $\xi $.

In summary, we have shown that the distribution of dipping with orbital
phase correlates approximately with the variation of column density 
of the neutral component of the stellar 
wind of HDE\thinspace 226868 with phase. There is in addition,
extra dipping at $\phi$ $\sim$ 0.6. These effects resemble the asymmetry
of absorption in the wind of Supergiant X-ray Binaries, suggesting
that the formation of absorbing blobs depends on the neutral density,
and thus reflects the $N_{\rm H}(\phi)$ variation. 

\section*{Acknowledgments}

This paper uses data made publicly available, and we
thank the ASM/{\it RXTE} Team, including members at MIT and NASA/GSFC. 
We thank Dr. J. Lochner for useful discussions on the flux/count 
rate calibrations for the ASM. MBC and MJC thank the British Council and the Royal
Society for financial support.

\end{document}